\documentclass[letterpaper,10 pt, conference]{ieeeconf}
\usepackage{color}
\usepackage{needspace}

\usepackage{theorem}
\usepackage{cite}
\usepackage{caption}
\usepackage{subcaption}

\usepackage{mathtools,amsmath,amssymb,amscd,verbatim}
\usepackage{xcolor}
\usepackage{multirow,bigdelim}
\usepackage{tikz}

\input{mysymbol.sty}

\newtheorem{theorem}{Theorem}

\newtheorem{algorithm}{Algorithm}
\newtheorem{proposition}{Proposition}

\newtheorem{lemma}{Lemma}

{\itshape}{\rmfamily}
\newtheorem{assumption}{Assumption}
\newtheorem{remark}{Remark}

\IEEEoverridecommandlockouts                              

\title{\LARGE \bf
Distributed Fictitious Play in Potential Games with Time-Varying Communication Networks
}

\author{Sina Arefizadeh and Ceyhun Eksin
\thanks{S. Arefizadeh and C. Eksin are with Department of Industrial and Systems Engineering,
        Texas A\&M University, College Station, TX 77845.
        {\tt\small eksinc@tamu.edu}}%
}

\begin{document}

\maketitle

\begin{abstract}
We propose a distributed algorithm for multiagent systems that aim to optimize a common objective when agents differ in their estimates of the objective-relevant state of the environment. Each agent keeps an estimate of the environment and a model of the behavior of other agents. The model of other agents' behavior assumes agents choose their actions randomly based on a stationary distribution determined by the empirical frequencies of past actions. At each step, each agent takes the action that maximizes its expectation of the common objective computed with respect to its estimate of the environment and its model of others. We propose a weighted averaging rule with non-doubly stochastic weights for agents to estimate the empirical frequency of past actions of all other agents by exchanging their estimates with their neighbors over a time-varying communication network. Under this averaging rule, we show agents' estimates converge to the actual empirical frequencies fast enough. 
This implies convergence of actions to a Nash equilibrium of the game with identical payoffs given by the expectation of the common objective with respect to an asymptotically agreed estimate of the state of the environment.
\end{abstract}

%
\section{Introduction} 

A networked multiagent system consists of heterogeneous agents that aspire to achieve a common objective by choosing their individual actions in the absence of a central coordinator. The common objective, which may represent a power control problem in wireless communications \cite{candogan2010near}, a distributed estimation problem \cite{chenSayed}, or a task given to a team of robots \cite{fink2013robust}, depends on an unknown environment variable in addition to the actions of \emph{all} agents. Here, we present a distributed algorithm for the scenario when agents disagree on their estimate of the environment, and thus of the objective. In such a setting unless agents wait or exchange information for multiple rounds, they cannot be sure about what other agents are optimizing. When information about the environment is streaming or the system is large-scale, waiting or communicating for multiple rounds before taking an action may be undesirable as it will incur long coordination delays. 

Here we propose a distributed algorithm for such scenarios where coordination delay is unreasonable. In the algorithm, agents keep an estimate of the environment and a model of how other agents' take their actions. Then, each agent \emph{best-responds}, i.e., takes the action that maximizes its expectation of the objective with respect to their estimate and model of behavior. The model of other agents' behavior assumes that each agent selects its actions from a stationary distribution given by the histogram of their past actions. This model is based on the fictitious play (FP) algorithm \cite{Brown_1951,Monderer_Shapley_1996a}. However, in a large-scale system, agents cannot observe the past actions of all the agents. Instead, here we consider a decentralized update scheme based on weighted averaging that allows agents to keep track of the histograms of all other agents when the communication network is time-varying. 

The proposed decentralized scheme generalizes prior work on distributed FP \cite{eksin2018distributed,Swenson_et_al_2014} to time-varying communication networks. We provide convergence rate of the decentralized weighted averaging updates to the true empirical frequencies when the weights matrix is row stochastic (see Proposition \ref{proposition_a}). Here, we build on distributed optimization algorithms that rely on reaching consensus fast enough \cite{tsitsiklis1984problems,NedicOzdaglar}. Unlike these prior works, we do not impose the weights of the averaging to be coordinated in order to satisfy a doubly stochastic assumption. The intuition behind our result is that each agent is \emph{stubborn} when it comes to keeping track of its own histogram of past actions, and other agents are following the stubborn agent's updates through information exchanges with their peers. As long as the time-varying network is connected over a union of past edges for some fixed finite-time, the stubborn agents' updates cascades down to the follower agents. 

Given the fast enough convergence of the estimates on others' empirical frequencies, and eventual agreement on the state of the environment, the distributed FP algorithm converges to the Nash equilibrium (NE) of the game where agents have identical payoffs computed by integrating the common objective with respect to the consensus estimates on the state of the environment. At an NE action profile, all agents act optimal with respect to the actions of other agents. Our convergence result relates to the literature on NE seeking algorithms \cite{Li_Basar_1987,Shamma_Arslan_2005,salehisadaghiani2016distributed,koshal2016distributed}. This work distinguishes from these NE seeking algorithms by not making any structural assumptions on the objective function, and considering unknown and time-varying payoffs due to evolving estimates of the environment. 
\section{Networked Multiagent Systems with Uncertainty} 

A group of agents $\ccalN = \{1, \ldots, n\}$ aims to maximize a common objective $u(a,\theta)$ that is a function of the joint action profile of all agents $a:=[a_1, \dots, a_n]$, and the state of the environment $\theta$ by selecting their individual actions $a_i$ belonging to a finite action space $A_i$. We define the space of joint action profile as $A = \prod_{i\in\ccalN} A_i$. The state of the environment $\theta$ is unknown. At subsequent points in time $t=0,1,\ldots$, agents simultaneously decide on an action $a_{i}(t)\in A_i$ that they deem optimal with respect to their current belief about the environment $\mu_{i}(t)$. Agent $i$'s belief about the environment $\mu_{i}(t)$ assigns probabilities to possible states of the environment $\Theta$, i.e., it belongs to the space of probability distributions over $\Theta$, denoted with $\Delta(\Theta)$.

If agents have different beliefs about the environment that is unknown to agent $i$, agent $i$ cannot be sure of the actions of other agents $a_{-i}(t):=\{a_{j}(t)\}_{j \in \ccalN\setminus i}$, hence it cannot be sure whether its action $a_{i}(t)\in A_i$ is optimal or not. In such a scenario, we assume agent $i$ keeps a belief about the choices of other agents $v^i_{-i}(t):=\{v^i_{j}(t)\}_{j\ne i}$ where $v^i_{j}(t)\in \Delta(A_j)$ is the belief of agent $i$ on agent $j$'s next action. Using its beliefs, agent $i$ takes the action that maximizes the expectation of the common objective computed with respect to its beliefs about the state and the actions of other agents,
\begin{align}
\label{eqD}
a_{i}(t)\in\underset{a_i\in {A_i}}{\argmax}\:u(a_i,v^i_{-i}(t);\mu_{i}(t))
\end{align}
where $u(a_i,v^i_{-i}(t);\mu_{i}(t))$ is the expectation of the objective with respect to the beliefs $v^i_{-i}(t)$ and $\mu_{i}(t)$.

\subsection{Communication}
Agents update their beliefs about the actions of other agents, $v^i_{-i}(t)$, by interacting with a subset of the agents in $\ccalN$. The subset of the agents that $i$ can interact with at time $t$ is determined by a network $\ccalG(t)$ with node set $\ccalN$ and a symmetric edge set $\ccalE(t)$. If the edge $(i,j)$ belongs to $\ccalE(t)$, agents $i$ and $j$ can exchange information with each other after decision epoch $t$. We denote the set of neighboring agents that interacts with $i$ at time $t$  as $\ccalN(i,t):=\{j: (i,j)\in \ccalE(t)\}$. We make the following assumption on the connectivity of time-varying networks.
\begin{assumption} \label{connectivity} The graph $(\ccalN,\ccalE(\infty))$ is connected where $(i,j)\in \ccalE(\infty)$ communicate infinitely many times, i.e., $\ccalE(\infty)=\{(i,j)| (i,j)\in \ccalE(t)$ for infinitely many $t\}$\cite{NedicOzdaglar}.
\end{assumption}

\begin{assumption} \label{BoundedIntercommunicationInterval}  There exist an integer $T\geq 1$ such that for every $(i,j)\in \ccalE(\infty)$ and $k\geq 0$, $(i,j)\in \ccalE(k)\cup\ccalE(k+1)\cup...\cup\ccalE(k+T-1)$. 
\cite{NedicOzdaglar}.
\end{assumption}

The Assumptions \ref{connectivity} and \ref{BoundedIntercommunicationInterval} made above are called 
\emph{connectivity} and \emph{Bounded intercommunication interval}, respectively in \cite{NedicOzdaglar}. Together connectivity and bounded intercommunication interval imply that the information generated agent $j\in\ccalN$ can reach agent $i\in\ccalN$ by some time. 

\begin{remark} \label{assump_1}  Define the network  $\ccalG(t,T):=(\ccalN,\ccalE(t,T))$ where $\ccalE(t,T):=\cup^{T-1}_{\tau=0} \ccalE(t+\tau)$ for $t>T$ where $t,T\in\naturals_+$. According to Assumption \ref{connectivity} and \ref{BoundedIntercommunicationInterval} $\ccalG(t,T)$ is strongly connected for each $t$.
\end{remark}


\subsection{Information Exchange and Belief Updates}

Agent $i$ assumes other agents are selecting their actions according to a stationary distribution, the empirical histogram of their past actions. The empirical histogram of agent $i$ at time $t$, denoted by $f_{i}(t)$, can be recursively updated as \cite{Monderer_Shapley_1996}
\begin{equation}
\label{eq1}
f_{i}(t+1)=f_{i}(t)+\frac{1}{t}(\Psi (a_{i}(t))-f_{i}(t)),
\end{equation}
where, $f_{i}(t)$ denotes the empirical histogram of agent $i$ at time $t$, and $\Psi(a_{i}(t))$ denotes an $|A_i|\times 1$ dimensional vector that is one at the $k$th element if $a_{i}(t)=k$ with $k\in A_i$, and otherwise it is zero. 

Agent $i$ cannot observe past actions of all the agents given the communication limitations. Hence, it is not possible for agent $i$ to keep track of the empirical histogram of other agents. Instead, agent $i$ will share and keep estimates of others' empirical frequencies in $v^i_{-i}(t)$. Specifically, at each step agent $i$ receives its current neighbors' estimates of agent $j$'s empirical frequency $\{v^{k}_{j}(t)\}_{k\in\mathcal{N}(i,t)\bigcup\{i\}}$ to update its estimate as follows,
\begin{equation}
\label{eq2}
v^i_{j}(t+1)=\sum_{k\in\mathcal{N}}w^i_{j,k}(t) v^{k}_{j}(t),
\end{equation}
where $w^i_{j,k}(t)$ denotes the weight that agent $i$ puts on $k$'s estimate of agent $j$ at time $t$. We make the following assumptions on the weights.
\begin{assumption}\label{assumption_weights}
Assume there exists a scalar $0<\eta<1$ such that for all $i\in \ccalN$, and $j\in\ccalN$,\\
{\it (i)} $w^i_{j,k}(t)\geq\eta$ only if $k\in\mathcal{N}(i,t)\bigcup\{i\}$, otherwise $w^i_{j,k}(t)=0$. \\
{\it (ii)} $w^i_{i,i}(t)=1$ for all $t$.\\
{\it (iii)} $\sum_{k \in\mathcal{N}}w^i_{j,k}(t)=1$ for all $t$.
\end{assumption}
We define the weights matrix $W_{j}(t)$ used for estimating agent $j$'s empirical frequency at time $t$, where element in the $i$th row and $k$th column of $W_j(t)$ is $[W_j(t)]_{i,k}=w^i_{j,k}(t)$, to discuss the implications of the above assumptions. Assumption \ref{assumption_weights}{\it (i)} makes sure that agents can only put positive weights on their current neighbors' estimates in \eqref{eq2}. Assumption \ref{assumption_weights}{\it (ii)} means that agent $i$ only listens to itself (\emph{stubborn}) when its empirical frequency is of concern, that is, we assume $v^i_{i}(t)= f_{i}(t)$. This implies that $j$th row of $W_j(t)$ is given by $\bbe_j^T$ which is an $1\times n$ row-vector of all zeros except 1 in the $j$th element. Assumption \ref{assumption_weights}{\it (ii)} also means that the weights matrix $W_j(t)$ is different for each agent $j$. Assumption \ref{assumption_weights}{\it (iii)} means that $W_j(t)$ is row-stochastic for all times. Note that we do not require $W_j(t)$ to be doubly-stochastic. In time-varying networks, requiring $W_j(t)$ to be column stochastic is unrealistic because it would necessitate agents to coordinate their weights at each step.

We will be agnostic to the individual updates on the state of the environment $\mu_{i}(t)$ as long as the state learning process satisfies the following assumption.

\begin{assumption}\label{assump_state}
The local beliefs on the state $\mu_{i}(t)$ converge to a common belief $\mu\in \Delta(\Theta)$ in terms of total variation,
\begin{equation}\label{eqn_state_belief_convergence} 
   \lim_{t\to\infty} {\bf TV}(\mu_{i}(t),\mu) = 0 \quad \forall i \in \ccalN,
\end{equation}
where the total variation distance between distributions $\mu_{i}(t)$  and $\mu$ is defined as the maximum absolute difference between the respective probabilities assigned to elements $B$ of the Borel set $\ccalB(\Theta)$ of the space $\Theta$, i.e., ${\bf TV}(\mu_{i}(t),\mu):= \sup_{B\in\ccalB(\Theta)} |\mu_{i}(t)(B) - \mu(B)|$. 
\end{assumption}
This assumption is equivalent to the one made in \cite{eksin2018distributed}. Next, we summarize the algorithm. 

\subsection{Decentralized Fictitious Play (D-FP) Algorithm}

\begin{algorithm}[D-FP algorithm]\label{algo_dfp_general} \normalfont
$~$\\
\noindent \textit{Initialize}\\
(i) For each $i$, let $a_{i}(0)$ be chosen arbitrarily, and let the estimate $v^i_{j}(1)$ be initialized as $f_{j}(1) = \Psi (a_{j}(1))$ for all $j\in \ccalN\setminus i$. Let $\mu_{i}(t)\in \Delta(\Theta)$ be arbitrarily chosen.
\medskip

\noindent \textit{Iterate} ($t\geq 1$)\\
(ii) Agents simultaneously choose their next-stage action according to the rule in \eqref{eqD}.

\noindent (iii) Agents update their empirical frequencies $f_{i}(t)$ as in \eqref{eq1}, and let $v^i_{i}(t) = f_{i}(t)$.

\noindent (iv) Each player $i$ engages in one round of information exchange with neighboring agents $j\in \ccalN(i,t)$ where they receive $v^j(t):=[f_{j}(t), v^j_{-j}(t)]$ and updates their estimate of the joint empirical distribution $v^i_{-i}(t)$ according to \eqref{eq2}.

\noindent (v) Agents update their beliefs about the environment $\mu_{i}(t)$ according to some state learning process.
\end{algorithm}

Step (ii) determines the actions, and steps (iii-v) determine how agents update their beliefs $v^i_{-i}(t)$ and $\mu_{i}(t)$. We assume that agents synchronously select their actions, and update their beliefs. However, the time-varying connectivity loosens this assumption to scenarios where some agents randomly wake up and send their beliefs to each other.

%
\section{Convergence}

We define the best individual action given the actions of others as the Nash equilibrium (NE) action profile---see Section \ref{sec_prelim} for a definition. We show convergence of the empirical frequencies of actions $f_{i}(t)$  generated by the D-FP algorithm converge to an NE (Theorem \ref{thm_NE}). The key technical contribution is in showing the convergence of beliefs $v^i_{j}(t)$ to true empirical frequency $f_{j}(t)$ with updates \eqref{eq2} at a fast enough rate given non-doubly stochastic weights  (Proposition \ref{proposition_a}). Given this convergence rate, the convergence to NE follows by results in \cite{eksin2018distributed}. Next, we introduce some preliminary technical concepts. 

\subsection{Preliminaries: Game Theory} \label{sec_prelim}

When the expectations of the objective are different, agents $\ccalN$ are playing a game $\Gamma$ with utility functions $u_{i,t}(\cdot):A \to \reals$, that is, $\Gamma=\{\ccalN, A, \{u_{i,t}\}_{i \in \ccalN}\}$. A mixed strategy $\sigma_i$ in a game corresponds to probability distribution over the action space $\Delta(A_i)$. We use $\sigma_i(a_i)$ to denote the probability that agent $i$ takes action $a_i\in A_i$. The joint mixed strategy profile is the product distribution of individual mixed strategies $\sigma:=\{\sigma_1,\sigma_2,...,\sigma_n\}$. We express the expected objective value with respect to the strategy profile $\sigma$ as
\begin{equation}
\label{eqA}
u(\sigma,\theta)=\sum_{a\in A} u(a,\theta)\sigma(a)
\end{equation}
where $A:=\prod_{i\in\ccalN} A_i$. We define agent $i$'s expectation of the common objective given its belief $\mu_{i}(t)$ about the environment as follows,
\begin{equation}\label{eq_utility}
u_{i,t}(\sigma):=u(\sigma; \mu_{i}(t)) = \int_{\theta \in \Theta} u(\sigma, \theta) \mu_{i}(t)(\theta)
\end{equation}
A (mixed) strategy profile $\sigma^*\in \prod_{i\in\ccalN}\Delta(A_i)$ is a Nash equilibrium of $\Gamma$ if no agent has unilaterally profitable deviation,
\begin{equation}
\label{eqC}
u_{i,t}(\sigma^*_i,\sigma^*_{-i})\geq u_{i,t}(\sigma_i,\sigma^*_{-i})\:\;  \forall \sigma_i \in \Delta(A_i).
\end{equation}
%

\subsection{Convergence of Beliefs on Empirical Frequencies}
Denote the vector that shows the estimation of the population on the frequency of agent $n$'s $l$th action as  $x(t):=[[v^1_{n}(t)]_l,\dots,[v^n_{n}(t)]_l]^T\in\mathbb{R}^{n\times 1}$, where the $k$th element of the vector is denoted by $x_k(t) = [v^k_{n}(t)]_l$. Recall that $[v^n_{n}(t)]_l= [f_n(t)]_l$ by Assumption \ref{assumption_weights}(ii). Thus $x_n(t)$ is updated according to the dynamics in \eqref{eq1}. Given the belief updates in \eqref{eq2}, we can write the linear dynamics for $x(t)$ as
\begin{equation} \label{eq_belief}
    x(t+1) = W(t) \big(x(t)+ (x_n(t+1)-x_n(t)) e_n\big)
\end{equation}
where $W(t)$ is the weights matrix for agent $n$ defined after Assumption \ref{assumption_weights} with subindex $n$ dropped, and $e_n$ is the $n$'s vector of the canonical basis in $\mathbb{R}^n$. The following result shows convergence rate of beliefs in \eqref{eq_belief} to true empirical frequency $f_{n}(t)$. 

\begin{proposition}\label{proposition_a} 
Let $x(t)\in \reals^{n\times 1}$ be a belief vector evolving according to \eqref{eq_belief} and the weights matrix $W(t)$ satisfying Assumption \ref{assumption_weights}. If the communication network satisfies Assumptions \ref{connectivity} and \ref{BoundedIntercommunicationInterval}, and $x_i(0)=x_n(0)$, then $||x_i(t)-x_n(t)|| = O(\frac{\log t}{t})$ for all $i\in \ccalN\setminus n$.
\end{proposition}
\begin{myproof}
Define $y(t) := x(t) - x_n(t) \bbone$ where $\bbone$ is a column vector of all ones. By subtracting $x_n(t) \bbone$ from both sides of \eqref{eq_belief}, we get
\begin{equation}
\label{eq5}
y(t+1)=W(t)(y(t)+\delta(t)),
\end{equation}
where $\delta(t):=(x_n(t+1)-x_n(t))(e_n-\bbone)$. Substituting  previous values of $y(s)$ for $s=0,\dots,t$ in \eqref{eq5}, we have
\begin{equation}\label{eq_recurion_y}
y(t+1)=\sum^{t-1}_{s=0}\big(\prod^{s}_{\tau=0} W(t-\tau)\big) \delta(t-s).
\end{equation}
where we used the assumption $x_i(0)=x_n(0)$ to get rid of the initial term containing $y(0)$. We take norms of both sides and bound the left hand side by moving the norm inside the summation
\begin{equation}\label{eq_bound_y}
||y(t+1)||\leq\sum^{t-1}_{s=0}||\big(\prod^{s}_{\tau=0} W(t-\tau)\big) \delta(t-s)||.
\end{equation}
Lemma \ref{lemma1} states that the products of weight matrices converge to $\bbone e_n^T$ with some rate $\rho$. Thus we can bound the right hand side above as follows,
\begin{equation}\label{eq_bound_y_3}
||y(t+1)||\leq \sum^{t-1}_{s=0}\rho^{s} ||\delta(t-s)||.
\end{equation}
%

%
Note that $\delta(t) \leq n/t$. Defining $\delta_{avg}(t):=\frac{1}{t}\sum^t_{s=1}\frac{n+1}{s}$, we can conclude 
$||y(t+1)||\leq\frac{\delta_{avg}(t)\rho}{1-\rho}$.
Result follows by noting that $\delta_{avg}(t)=O(\frac{log t}{t})$. 
\end{myproof}

The result shows that agents are able successfully track estimates of an arbitrarily selected agent $n$. When agents are able correctly estimate the empirical frequencies of other agents, the algorithm is close to a centralized FP algorithm from which convergence to NE follows as we state next. 

\begin{theorem} \label{thm_NE}
Let Assumptions \ref{connectivity}, \ref{BoundedIntercommunicationInterval}, \ref{assumption_weights}, and \ref{assump_state} hold. Define the game with common state belief $\mu$ and identical payoffs $u_{i,\infty}$ as $\Gamma(\mu)$. The empirical frequency of actions generated by Algorithm \ref{algo_dfp_general} converge to a NE strategy of $\Gamma(\mu)$, 
\begin{align} \label{histogram_convergence}
 \lim_{t\to\infty} \min_{\sigma^* \in K(\mu)} \| f_t - \sigma^*\| = 0
\end{align}
where $K(\mu)$ represents the set of Nash equilibria of $\Gamma(\mu)$, i.e., all $\sigma^*$ that satisfy \eqref{eqC}. 
\end{theorem}

Proof of the above result follows by first observing that $\Gamma(\mu)$ is an identical interest potential game \cite{Monderer_Shapley_1996}. Second, we observe that Proposition \ref{proposition_a} satisfies the same convergence rate as its counterpart (Lemma 1) in \cite{eksin2018distributed} for fixed connected communication networks. Thus the proof of Theorem \ref{thm_NE} follows verbatim the proof of Theorem 1 in \cite{eksin2018distributed}. 
\section{Simulation}

$n=5$ agents are tasked with covering $n$ targets. The global objective is given as 
\begin{equation} \label{eq_task_assignment}
    u(a,\theta)=\sum_{i=1}^n 1\bigg(\sum_{j\ne i}1(a_j=k)=0 \bigg)||x_i-\theta_k||^{-2}
\end{equation}
where $x_i$ and $\theta_k$ are the locations of the agent $i$ and target $k$, respectively. As per \eqref{eq_task_assignment}, agents receive a zero payoff from a target if more than one agent is covering it. The payoff agent $i$ can receive from selecting a target is inversely proportional to its distance to the target. Target locations are unknown. Agents receive private noisy signals about target locations at each step. In the target assignment game with common utility function in \eqref{eq_task_assignment} and common beliefs on the state, there are multiple Nash equilibria. In particular, any action profile that covers all targets is a NE. 

\begin{figure}
      \centering
\begin{tabular}{cc}
        \includegraphics[width=.47\linewidth]
        {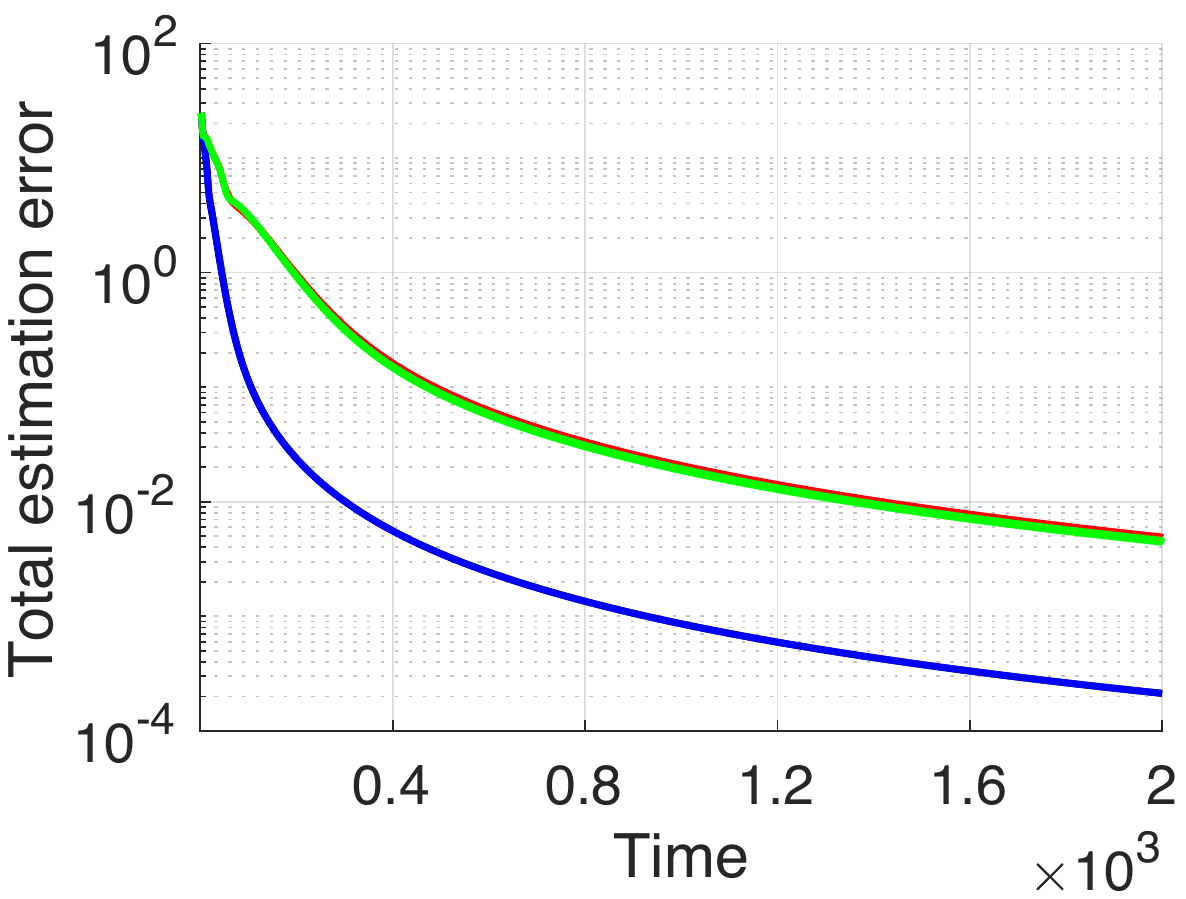}&
        \includegraphics[width=.47\linewidth]
        {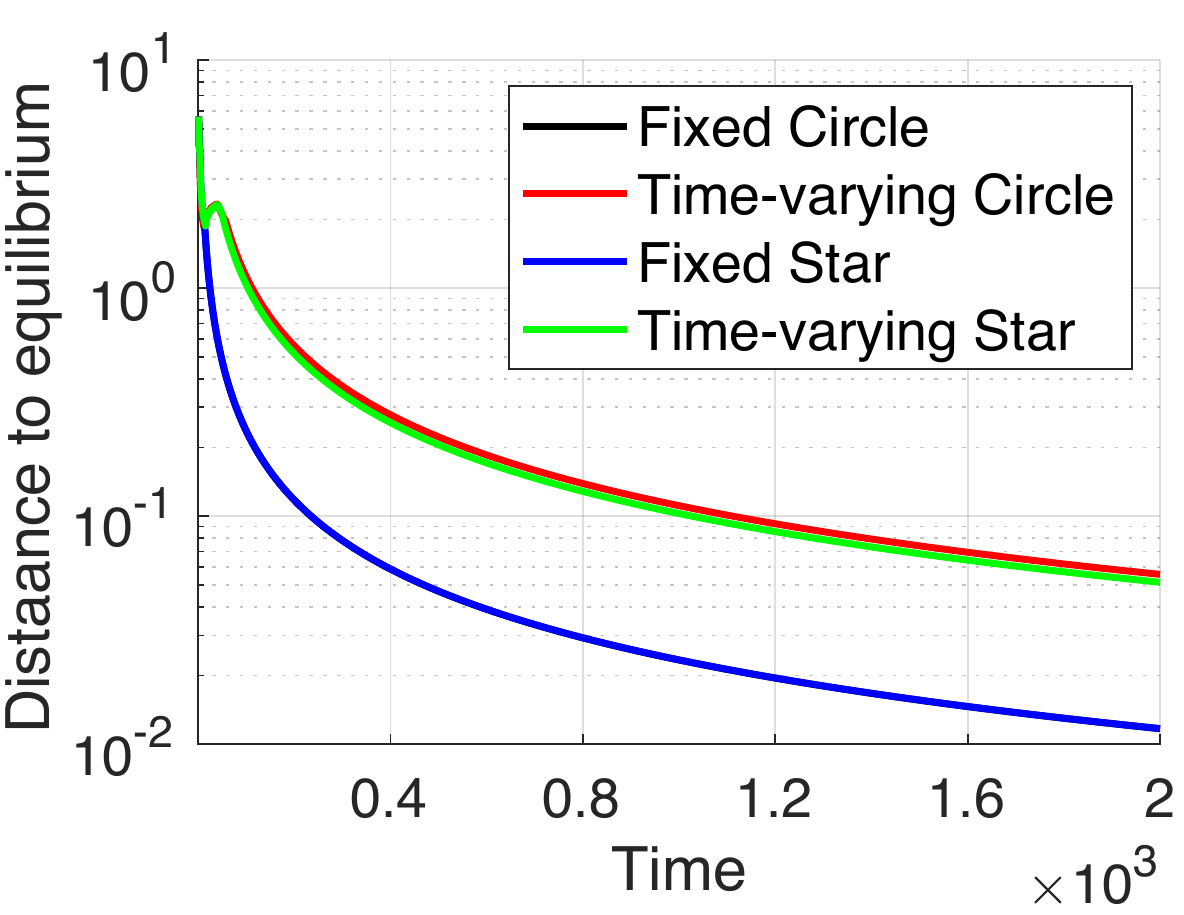}
        \end{tabular}
        \caption{(Left) Convergence of estimates $\sum_{j\in\ccalN}\sum_{i\in\ccalN}||\hat v^i_j(t)-f_j(t)||$ (Right) Convergence of empirical frequencies to NE ($\sum_{i\in\ccalN}||f_{i}(t)-\sigma^*||$) where $\sigma^*$ is an NE strategy.}
              \label{fig3}
      \end{figure}

Fig. \ref{fig3} compares convergence rates for fixed and time-varying communication networks (ring and star). In the time-varying networks each edge in the star (or ring) network appears one at a time similar to gossiping schemes satisfying Assumption \ref{assump_1} for $T=5$. Fig. \ref{fig3}(Left) shows that total error on estimates of empirical frequencies converges at the same rate when the network is time-varying as when the network is fixed. This plot confirms $O(\frac{\log t}{t})$ rate shown in Proposition \ref{proposition_a}. Fig. \ref{fig3}(Right) shows the rate of convergence to a NE action profile $\sigma^*$ for the run considered. While convergence of empirical frequencies to $\sigma^*$ is as shown, agents start acting according to the NE action profile $\sigma^*$, i.e., each agent selects a different target, after $t=50$ and $t=57$ for time-varying star and ring networks, respectively.

\section{Conclusion}

In this paper, we proposed a variant of the distributed fictitious play for time-varying communication networks. In the algorithm, agents keep estimates of empirical frequency of others' actions by sharing their estimates with their current neighbors and updating their estimates using weighted averaging. We showed that convergence rate of the estimates are fast enough to guarantee convergence of the empirical frequencies of actions to an NE of the game, where eventually agents have identical expectations of the common objective. The key technical novelty is that the weights matrix is only row (non-doubly) stochastic which means that there is no need for coordination of weights.

\bibliographystyle{IEEEtran}
\bibliography{bibliography}

\begin{appendix}

Define $\Phi(t,s) := W(t)W(t-1)\cdot\cdot\cdot W(s)$ for $s\leq t$ where $W(t)$ is the weights matrix in \eqref{eq_belief} for keeping track of agent $n$'s empirical frequency. We overload notation to define $W(t,k):=\Phi(t+ k T - 1, t+(k-1)T)$ for $k\in\naturals_+$. Note that the sparsity of $W(t,k)$ corresponds to a strongly connected network as per Assumptions \ref{connectivity} and \ref{BoundedIntercommunicationInterval}. $W(t,1:d):=W(t,d)W(t,d-1)\cdot\cdot\cdot W(t,1)$. For an arbitrary matrix $W$, we denote its element in the $i$th row and $j$th column with $[W]_{i,j}$. The matrix with $n$th column and $n$th row removed is denoted with $[W]_{1:n-1,1:n-1}$. 

\begin{lemma} \label{lemma1} Let Assumptions \ref{connectivity}, \ref{BoundedIntercommunicationInterval} and \ref{assumption_weights} hold. Then we have\\
{\bf (a)} $\lim_{t\to\infty}\Phi(t,s)  = \bbone e_n^T$ for any $s\in \naturals_+$.\\
{\bf (b)} $|[\Phi(t,s)]_{i,j} - [e_n]_j|\leq \kappa \rho^{t-s}$ for some $0<\rho<1$ and $\kappa>0$.
\end{lemma}
\begin{myproof}
Let $t=l+kT-1$ for given $k,s \in \naturals_+$ with $l>s$. we can write
\begin{equation*}
    \phi(l+kT-1,s)=\phi(l+kT-1,l)\cdot\phi(l-1,s).
\end{equation*}
Thus we can write 
\begin{equation}
\label{k_multiplication}
    \phi(l+kT-1,s)=[\prod^k_{r=1}W(l,r)]\cdot\phi(l-1,s).
\end{equation}
Let $d=(n-1)T$. By Lemma \ref{lem2 of Nedic's paper}, putting $s$ value to be $l+md+1$ the last column of the matrix product $W(l,md+1:(m+1)d)$ for every $m$ is a positive vector. Let $k=md$ in \eqref{k_multiplication}, then we have
\begin{equation}
\label{md_multiplication}
    \phi(l+mdT-1,s)=[\prod^{md}_{r=1}W(l,r)]\cdot\phi(l-1,s)
\end{equation}
\begin{equation*}
 =W(l,1:md)\cdot\phi(l-1,s).
\end{equation*}
For $l>s$, $\phi(l-1,s)$ is row stochastic and $[\phi(l-1,s)]_{n,1:n}=e^T_n$. All columns of $\phi(l-1,s)$ except for $n$-th column satisfy condition of $x_0$ and $W(l,rd+1:(r+1)d)$ satisfy condition for matrix $D_r$ for each $r$ in Lemma \ref{geometrically convergance}, then 
\begin{align}
 [W(l,&1:md)\cdot\phi(l-1,s)]_{i,j\neq n} \nonumber\\ \leq &(1-\eta^{(n-1)T})^m||\phi(l-1,s)]_{1:n,j\neq n}||_\infty.
\end{align}
Since $||\phi(l-1,s)]_{1:n,j\neq n}||_\infty \leq 1$ we have 
\begin{equation}
\label{phi_upperbound}
[\phi(l+mdT-1,s)]_{i,j\neq n}\leq (1-\eta^{(n-1)T})^m. \: \forall t>s
\end{equation}
Considering $l+mdT-1=t$ and $s\leq l-1<s+dT$, we have $s+mdT\leq t<(m+1)dT$. As a result, we can write
\begin{align}
(1-\eta^{(n-1)T})^m&=\frac{(1-\eta^{(n-1)T})^{m+1}}{(1-\eta^{(n-1)T})}\\
&=\frac{(1-\eta^{(n-1)T})^{\frac{s+(m+1)dT-s}{dT}}}{(1-\eta^{(n-1)T})}\\
&\leq\frac{(1-\eta^{(n-1)T})^{\frac{t-s}{dT}}}{(1-\eta^{(n-1)T})}.
\end{align}
Thus, for every $t,s$ with $t\geq s$ we can write 
\begin{equation}
|[\phi(t,s)]_{i,j\neq n}-0|\leq \frac{\kappa}{n-1}\rho^{t-s}
\end{equation}
where $\kappa = \frac{(n-1)}{(1-\eta^{(n-1)T})}$ and $\rho=(1-\eta^{(n-1)T})^{\frac{1}{dT}}$.
Because the matrix $\Phi(t,s)$ is row stochastic,
$\left|[\phi(t,s)]_{i,n}- 1\right|\leq\kappa \rho^{t-s}$. 
Parts {\bf (a)} and {\bf (b)} follow from above. 
\end{myproof}

The proof above follows similar steps as in Lemmas 1-4 in \cite{NedicOzdaglar}. The difference here is that we show the limiting matrix is a specific rank one row-stochastic matrix, i.e., $\bbone e_n^T$ while in \cite{NedicOzdaglar} it is shown that the limiting matrix is $\bbone \xi(t)^T$ where $\xi(t)$ is a stochastic vector. The key difference that leads convergence to the specific stochastic vector is that agent $n$ only puts weight on itself (Assumption \ref{assumption_weights}(ii)) while in \cite{NedicOzdaglar} this assumption is not made. Note that  Assumptions \ref{assumption_weights}(i) and (iii) is equivalent to Assumption 1 in \cite{NedicOzdaglar}. In a sense agent $n$ is stubborn when it comes to its own empirical frequency which leads to other agents \emph{following} agent $n$'s updates. 

\subsection{Technical Results}

\begin{lemma} \label{lem1 of Nedic's paper} Let weight rule of assumption \ref{assumption_weights}{\it (i)} holds true then\\
{\bf (a)} $[\phi(t,s)]_{j,j} \geq \eta^{t-s+1}$ for all $j,t,s$, with $t\geq s$\\
{\bf (b)} $[\phi(t,s)]_{i,j} \geq \eta^{t-s+1}$ for all $t,s$ with $t\geq s$ for all $i\neq n$ and $j$, where  $(i,j)\in \ccalE(s)\cup\ccalE(s+1)\cup ...\cup\ccalE(t)$.\\
{\bf (c)} Let $i\neq n$ and $v\neq n$, $(i,v)\in \ccalE(s)\cup\ccalE(s+1)\cup ...\cup\ccalE(r)$ for some $r\geq s$ and $(v,j) \in \ccalE(r+1)\cup\ccalE(r+2)\cup ...\cup\ccalE(t)$ for $t>r$. Then $[\phi(t,s)]_{i,j} \geq \eta^{t-s+1}$.\\
{\bf (d)} $\phi(t,s)$ for each $t,s$ with $t\geq s$ is row stochastic. 
\end{lemma}
\begin{myproof}
The proof is similar to proof of lemma 1 of \cite{NedicOzdaglar}. 
\end{myproof}

\begin{lemma} \label{lem2 of Nedic's paper} Let Assumption \ref{connectivity}, \ref{BoundedIntercommunicationInterval}, and \ref{assumption_weights}, hold. For all $s$, $i\neq n$ we then have $[\phi(s+(n-1)T-1,s)]_{i,n}\geq \eta^{(n-1)T}$.
\end{lemma}
\begin{myproof}
The proof is similar to proof of lemma 2 of \cite{NedicOzdaglar} using lemma \ref{lem1 of Nedic's paper} of this paper instead of lemma 1 of \cite{NedicOzdaglar}. The main difference is that we only consider paths that end at $n$ to make a claim about the last column of product. 
\end{myproof}

\begin{lemma} \label{geometrically convergance} Let $x_0$ be a column vector of size $n$ such that $[x_0]_n=0$ and $[x_0]_i$ is arbitrary for $i\neq n$. Let $D_k$ for every $k$ be squared row stochastic matrix of size $n$ where its last column is a positive vector with elements greater than or equal to some $\zeta$ and $[D_k]_n=e^T_n$. Then if $x_{k+1}=D_k\cdot x_k$ for every $k$, we have 
\begin{equation}
    [x_{k+1}]_i\leq (1-\zeta)^k||x_0||_\infty \: \forall i\neq n 
\end{equation}
\end{lemma}
\begin{myproof}
We know $[x_{k+1}]_n=[D_{k}]_n\cdot x_k=0$. Hence
\begin{equation}
    [x_{k+1}]_i=[D_k]_i\cdot x_k= \sum^{n-1}_{j=1}[D_k]_{i,j}[x_k]_j+0.
\end{equation}
Since $D_k$ is row stochastic, we have $\sum^{n-1}_{j=1}[D_k]_{i,j}=1-[D_k]_{i,n}\leq 1-\zeta$. Thus
\begin{equation}
    [x_{k+1}]_i\leq (1-\zeta)\cdot||x_k||_\infty \: \forall i\neq n
\end{equation}
As a result
\begin{equation}
    [x_{k+1}]_i\leq (1-\zeta)^k\cdot||x_0||_\infty \: \forall i\neq n
\end{equation}

\end{myproof}

\end{appendix}

\end{document}